\documentclass[conference]{IEEEtran}
\IEEEoverridecommandlockouts

\makeatletter
\def\markboth#1#2{\def\leftmark{\@IEEEcompsoconly{\sffamily}\MakeUppercase{\protect#1}}%
\def\rightmark{\@IEEEcompsoconly{\sffamily}\MakeUppercase{\protect#2}}}
\makeatother

\usepackage[utf8x]{inputenc}
\usepackage[english]{babel}
\usepackage{ucs}
\usepackage{amsmath}
\usepackage{amsfonts}
\usepackage{amssymb}
\usepackage{amsthm}
\usepackage{array}
\usepackage{verbatim}
\usepackage{listings}
\usepackage{stfloats}

\usepackage{algorithm}
\usepackage{algpseudocode}
\usepackage{url}
\usepackage{enumerate}
\usepackage{multirow}

\usepackage{epsfig}
\usepackage{epstopdf}
\usepackage{graphicx}
\usepackage{caption}
\usepackage{subcaption}

\def\beq{\begin{equation}}
\def\eeq{\end{equation}}
\def\beqa{\begin{eqnarray}}
\def\eeqa{\end{eqnarray}}
\def\beqan{\begin{eqnarray*}}
\def\eeqan{\end{eqnarray*}}

\title{Dynamic Time-domain Duplexing for Self-backhauled Millimeter Wave Cellular Networks}

\author{
    \IEEEauthorblockN{
	    Russell Ford~\IEEEmembership{Student Member,~IEEE},
	    Felipe G\'omez-Cuba~\IEEEmembership{Student Member,~IEEE}
	    }
	\IEEEauthorblockN{
		Marco Mezzavilla~\IEEEmembership{Student Member,~IEEE},
	    Sundeep Rangan,~\IEEEmembership{Senior Member,~IEEE}
	    }
}

\begin{document}
\maketitle

%\begin{abstract}
%
%\end{abstract}

%\begin{IEEEkeywords}
%Wideband regime, capacity scaling laws,
%\end{IEEEkeywords}

\begin{abstract}
Millimeter wave (mmW) bands between 30 and 300~GHz have attracted considerable
attention for next-generation cellular networks due to vast quantities of
available spectrum and the possibility of very high-dimensional antenna arrays.
However, a key issue in these systems is range: mmW signals are extremely vulnerable
to shadowing and poor high-frequency propagation.  Multi-hop relaying is therefore a natural
technology for such systems to improve cell range and cell edge rates without the addition
of wired access points.
This paper studies the
problem of scheduling for a simple infrastructure cellular relay system
where communication between wired base stations and User Equipment follow a hierarchical tree structure
through fixed relay nodes. Such a systems builds naturally on existing cellular
mmW backhaul by adding mmW in the access links.  A key feature of the proposed system
is that TDD duplexing selections can be made on a link-by-link basis due to directional isolation from other links.
We  devise an efficient, greedy algorithm for centralized scheduling that maximizes network utility by jointly optimizing the duplexing
schedule and resources allocation for dense, relay-enhanced OFDMA/TDD mmW networks. The proposed algorithm
can dynamically adapt to loading, channel
conditions and traffic demands.
%and can easily incorporate  additional constraints such as minimum delay for individual flows or classes of data and UE energy consumption.
Significant throughput gains and improved resource utilization offered by our algorithm over the static, globally-synchronized TDD patterns are demonstrated through simulations based on empirically-derived channel models at 28~GHz.
\end{abstract}

\section{Introduction}
Millimeter wave (mmW) networks are an attractive candidate for beyond 4G and 5G
cellular system evolution.  Such systems can potentially offer
tremendous increases in bandwidth
along with further gains from highly directional antenna arrays \cite{KhanPi:11-CommMag,RanRapE:14}.
However, a key issue in these systems is cell coverage and range due
to the extreme susceptibility of these high-frequency signals
to shadowing and high isotropic propagation loss.
Given these range limitations,
multi-hop relaying is a natural technology in the mmW space
\cite{taori2014band,kim2014system}.
Furthermore, multi-hop relaying is particularly attractive for mmW systems since many cellular
systems already use mmW backhaul and thus the addition of access links
will naturally give rise to multi-hop systems.

Most mmW systems designs have assumed a time-division duplex (TDD)
structure to fully exploit beamforming and eliminate the need for
paired bands.  In current TDD cellular standards, such as TD-LTE,
all subframes are globally synchronized with base stations transmitting in one common set of DL time slots and User Equipment (UEs) in transmitting in the complementary UL set~\cite{3GPP36.300}.
The DL/UL transmission mode patterns are essentially static and cannot be adjusted for load balancing or changing channel conditions experienced by mobile nodes.
Additionally, in the current LTE Release 10 specifications~\cite{3GPP36216}, in-band relay communication between the ``donor" eNodeB and Relay Node (RN) can occur only in designated subframes, which, as we will show in Section \ref{sec:sim_setup}, could result in the wireless backhaul being severely bottlenecked.
In the case of such traditional relay-enhanced networks, specific subframes must be reserved for the purpose of Inter-cell Interference Coordination (ICIC): Since eNBs typically transmit at a much greater power than the RNs, if these nodes transmit together, the RN signals may be overwhelmed by interference from the donor cell.

Such static, synchronized duplexing may not be
necessary and, in fact, may be particularly disadvantageous for mmW systems and wireless systems that use high-gain, directional antennas.
In these systems, interference from transmitters can be isolated even if there are significant power disparities (as found in \cite{AkdenizCapacity:14}).
In this work, we thus consider a \emph{dynamic duplexing} scheme
where individual links can select their own transmit-receive
duplexing pattern.
Specifically, we consider a system where subframes are synchronized network-wide in time,
but the transmit/receive selections in each subframe
can be made on a link-by-link basis -- a feature uniquely available
in the mmW range due to directional isolation.
This flexibility
enables the duplexing pattern to be dynamically optimized to current
traffic loading and channel conditions.
In addition, the duplexing can be
adapted to local topological constraints.  This adaptation is particularly
valuable since the number of hops and their capacity are
likely to vary significantly due to different cell sizes, propagation
obstacles and availability and quality of wired backhaul.

\paragraph*{Related work}
There is now a large body of work in
optimization, scheduling, power control and relay selection in OFDMA/TDD cellular networks \cite{HuangL:07,WangL:10,AlRawi:11,Ruang:12,Ahmed:12,ZhangX:13}, including several that propose dynamic TDD algorithms designed to take
advantage of the new LTE-B
enhanced Interference Mitigation and Traffic Adaptation (eIMTA) capabilities \cite{YoB:12,Huang:14,3GPP36828}.
These works focus mainly on scheduling and ICIC in the context of 4G microwave networks under the assumption of an interference-limited regime, whereas we assume constant interference due to directional isolation.
We also take the approach of centralized scheduling, which is in contrast to works on distributed MAC schemes for mesh networks, as in \cite{mudumbai2009}.
Also, while our analysis is based on simulation, we point out that
stochastic geometry analysis of self-backhaul mmW networks has recently
appeared in \cite{Singh:14arxiv} as well as a scaling law analysis
in \cite{gomez2014scaling}.

\section{System Model}
\label{sec:system_model}
~
\begin{figure}[!ht]
\centering
\includegraphics[width=2.3in,trim=.5in .3in .5in .5in]{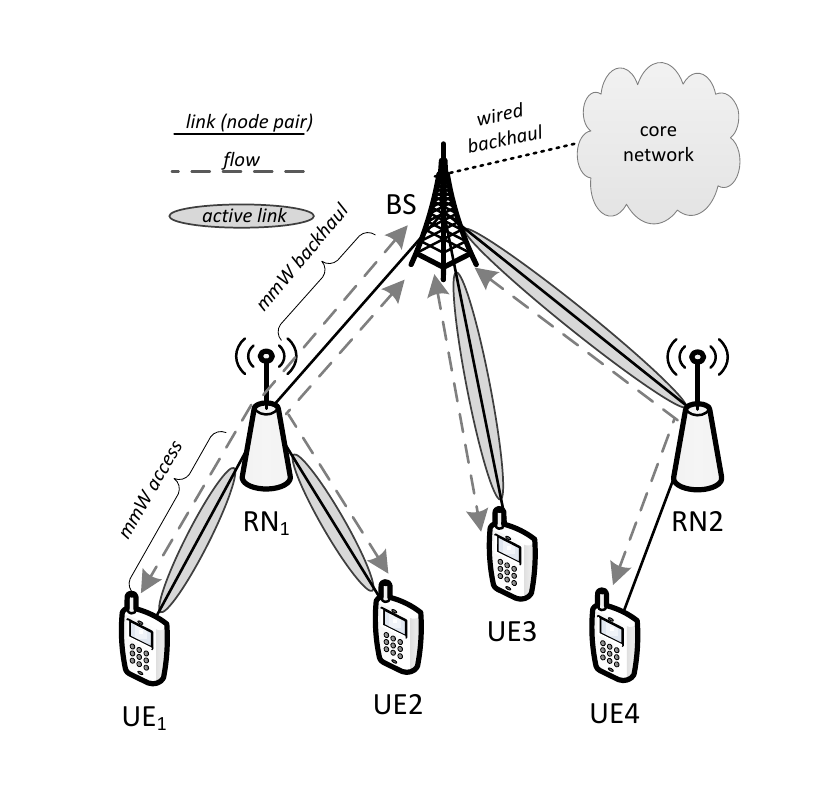}
\caption{Directional multi-hop cellular network with millimeter wave backhaul and access links}
\label{fig:network}
\end{figure}

\subsection{Network Topology}
We consider an OFDMA-TDD cellular network with directional smart antennas and multi-hop, in-band relaying as shown in Fig.~\ref{fig:network}. 
We assume a tree-structured network, where an eNodeB base station (denoted $BS$)
with wireline backhaul provides a root from 
which connections go to $N_{UE}$ user equipment
mobiles ($UE_1,\ldots,UE_{N_{UE}}$) via $N_{RN}$ relay nodes 
$RN_1\dots RN_{N_{RN}}$ that are self-backhauled, decode-and-forward stations.
RNs are essentially indistinguishable in operation from the BS but utilize wireless backhaul to the primary BS, which, unlike the BS's wired backhaul connection, is subject to the impairments of the mmW channel. 
For exposition, we assume a two-hop network, although all the methods
apply to multihop networks as well.
Each user is associated with either the BS (i.e. direct-link) or a relay $RN_j$. The rate on the link between $RN_j$ and its parent BS is denoted $R_{l_{1,j}}$ and $R_{l_{j,1}}$ for the downlink (DL) and uplink (UL), respectively, The rates on the DL and UL links between $UE_i$ and its parent BS or RN are likewise written $R_{l_{j,i}}$ and $R_{l_{i,j}}$.

We define $\mathcal{F}$ to be the set of flows in the network. Users have one DL flow and one UL flow, which gives $N_{\mathcal{F}}=2N_{UE}$ total flows in $\mathcal{F}$, but it can be easily extended to allow for multiple differentiated types of traffic. Flows have a throughput $R_{f}$ and utility $U_f(R_{f})$.
We assume a standard  
proportional-fair metric \cite{KelleyMT:98}
\begin{equation}
\label{eqn:utility}
U_f(R_{f}) = \log(R_{f}),
\end{equation}
although other concave utilities may also be used.

Users are scheduled in a series of epochs or frames of period $T_f$, which are further subdivided into $N_{sf}$ subframes of period $T_{sf} = T_f / N_{sf}$. 
From the perspective of each node, each subframe 
can be designated for DL or UL transmission indicating whether the transmissions
in that subframe are away from or toward the BS root of the tree.
A subframe can also be muted (i.e. unutilized).
Within each subframe, we assume Orthogonal Frequency Division Multiple Access (OFDMA) is employed, allowing multiple users to be allocated orthogonal frequency resources within same subframe. Specific OFDMA parameters, discussed in Section \ref{sec:sim_setup}, are derived from the LTE-like mmW system design proposed in \cite{PiSysDes:11}.

In contrast to the semi-static and globally synchronized TDD configurations supported by relay-enhanced TD-LTE networks, we allow each individual BS and relay node to dynamically select the transmission mode of each subframe. The assignment is centrally-coordinated through control messaging from the BS, however the operation of the specific MAC protocol is beyond the scope of this work.

Additionally, we make the following assumptions about resource assignment and the network, in general:
\begin{enumerate}[i.]
\item{\textit{Half-duplex --}\label{hd_assump}} Transmission and reception cannot occur simultaneously (i.e. during the same subframe) between pairs of adjacent nodes.

\item{\textit{Constant-interference --}} Wireless links behave nearly like point-to-point links due to the spatial isolation of directional beams. This is the key insight that allows us to analyze each routing tree, rooted at the BS, separately, and schedule resources without need for interference coordination. Based on the findings in
\cite{AkdenizCapacity:14}, 
we assume that high-gain, directional beamforming, enabled by multi-element antenna arrays, provides for minimal interference. However, receiving nodes may experience some small but non-negligible interference power.  To be conservative,
the interference at each node is taken to be the power assuming 
all nodes in the network were to be simultaneously transmitting at full power, with averaged TX/RX beamforming directions. 

\item{\textit{Single-stream} --} \label{ss-assumption} A node can transmit to only one receiver in the same time slot and frequency resources. In later work, we shall consider BS and RN nodes capable of multi-user Space Division Multiple Access (SDMA), which can more efficiently utilize the degrees of freedom of the channel than standard OFDMA \cite{TseV:07}. %\cite{TseV:07,KarLS:08,MedlesA:06}

\item{\textit{Constant-channel} --} As in \cite{AkdenizCapacity:14}, 
we assume that shadowing and other large-scale channel parameters vary slower than the duration being simulated. Since resource allocation is only modeled over a small sequence of frames, these stochastic parameters remain constant for the entire sequence. Small-scale fading is accounted for in our formulation of achievable rate.

\item{\textit{Single-path and static routing} --} \label{fixed-topo-assump} Each RN has a single link back to the BS and UEs are associated only with the BS or RN that offers the least path loss. The topology of the network is therefore fixed over the entire sequence of frames. Furthermore, it is assumed that relays are deployed by the operator as to guarantee a minimum capacity for the mmW backhaul link between the RN and wired BS, meaning that UEs will not associate an RN that cannot provide a quality backhaul connection. 
%Future work shall study the more complex problem of joint user-association, routing and resource allocation.
\end{enumerate}

\subsection{Channel Model}
For the simulation, 
we adopt the distance-based path loss model for a dense, urban environment at 28~GHz developed in \cite{AkdenizCapacity:14}, which itself is empirically derived from extensive measurements taken in New York City \cite{Rappaport:12-28G,Samimi:AoAD}.
Each link from an access point (BS or RN) to a UE are randomly in
one of three states:
non-line-of-sight (NLOS), line-of-sight (LOS) or outage
(i.e. the signal is blocked completely by some obstruction in the environment).
For nodes with a NLOS 
path to their selected transmitter, i.e. the serving BS or RN, the path loss is represented by (\ref{eqn:pathloss}),
\begin{equation} \label{eqn:pathloss}
PL_{dB}(d) = \alpha + \beta 10 \log 10(d) + \xi\quad \xi \sim \mathcal{N}(0,\sigma^2),
\end{equation}
where $d$ is in meters and the parameters 
$\alpha$, $\beta$ and $\sigma$ are provided in table \ref{tab:sim_params}
-- see \cite{AkdenizCapacity:14} for details.

As earlier noted, UE nodes initially select their serving BS or RN node with the least path loss. The \textit{long-term} beamforming gain is then applied, which is a function of the expected gain given the optimal TX and RX beamforming vectors, which, in turn, depends on the number of antenna elements ($n_{TX}$ and $n_{RX}$) in the $\lambda / 2$ planar antenna array. We again defer to \cite{AkdenizCapacity:14} for further details on how the long-term beamforming gain is calculated, as this subject is outside the scope of this paper.

\subsection{SINR and Rate Calculation}
As discussed in the Introduction, for the purpose of the optimization,
we assume a conservative interference model where the 
Signal-to-Interference-Noise Ratio (SINR) from any transmitting 
node $i$  to a receiving node $j$ is computed as 
\begin{align}  \label{eqn:sinr}
    \mbox{SINR}_{ij} =  \frac{P_{RX,ij}}{P_{int,max,i} + P_{N,i}},
\end{align}
where $P_{RX,ij}$ is the received power from node $i$ to node $j$
assuming the path loss and 
optimal long-term beamforming gain; $P_{int,max,i}$ is
the maximum average interference and $P_{N,i}$ is the thermal noise.
The maximum average interference is computed assuming that (a) all other
nodes are transmitting and (b) each of the other nodes 
select a beamforming direction to a random node that it could possibly 
transmit to. When a node such as a BS or RN transmits to multiple
other nodes, the transmit power is scaled by the bandwidth allocated
to each of the receivers.
Given the SINR and bandwidth allocation, we assume that 
maximum data rate for (DL or UL) link $l_{j,i}$ is calculated based on the formulation in \cite{MogEtAl:07}, which abstracts the achievable capacity
for LTE/LTE-A systems as
\begin{equation} \label{eqn:rate}
C_{ij} = \eta W_{ij} \log_2(1 + 10^{0.1 (\mbox{SINR}_{ij} - \Delta_{loss})}) ,
\end{equation}
where $W_{ij}$ is the bandwidth allocated to the link, $\eta$ is the bandwidth overhead
and $\Delta_{loss}$
accounts from loss from Shannon capacity due to 
small-scale fading, channel coding and inaccuracy of Channel State Information (CSI).  Following \cite{AkdenizCapacity:14}, we take $\eta = 0.8$ and $\Delta_{loss} = $ 3 dB.

\section{Optimization Formulation}
\label{sec:opt_problem}
We consider the problem of maximizing the sum utility,
\beq \label{eq:Uopt}
    U(\overline{R}) := \sum_f U_f(R_f),
\eeq
where $\overline{R}$ is the vector of the flow rates $R_f$, $f \in {\mathcal F}$.
We divide the frame into $N_{\rm sf}$ subframes, and 
in each subframe, $t=1,\ldots,N_{\rm sf}$,
and each link $\ell$, we let $x^t_\ell \in \{0,1\}$ be the binary
variable indicating whether there is a transmission on that link in that subframe.
Determining the binary variables $x^t_\ell$, in effect, determines
the TDD scheduling pattern.  To enforce the half-duplex constraint, we require that in each subframe $t$,
\beqa
    & x_\ell^t + x_{\ell'}^t \leq 1, \forall \ell=(p,n),~ \ell'=(n,c), \label{eq:conHD1} \\
    & x_\ell^t + x_{\ell'}^t \leq 1, \forall \ell=(n,p),~ \ell'=(c,n), 
    \label{eq:conHD2}
\eeqa
for all links $\ell=(p,n)$ and $(n,p)$ between a node $n$ and its parent $p$ 
and $\ell'=(n,c)$ and $(c,n)$, which the links from $n$ to child $c$.
In addition to the transmission schedule variables,
the optimization will also allocate bandwidths $W_\ell^t$ for each link $\ell$
in subframe $t$ with a maximum bandwidth constraint
\beq \label{eq:Wcon}
    \sum_{\ell \in {\mathcal L}(n)} W_\ell^t \leq W_{max},
\eeq
where $W_{max}$ is the total available bandwidth and ${\mathcal L}(n)$
are the set of links transmitted from node $n$.  
We additionally require that $W_\ell^t = 0$ when $x_\ell^t=0$.
Given the bandwidth allocations $W_{\ell}^t$, determines the link capacity
$C_\ell^t$ from \eqref{eqn:rate}.
We let $R_{f,\ell}^t$ be the rate allocated to flow $f$ on link $\ell$ in
subframe $t$ which must satisfy the flow and capacity constraints
\beqa
    & \sum_{f \in {\mathcal F}(\ell)} R_{f,\ell}^t \leq C_\ell, 
       \label{eq:capCon} \\
    & R_f \leq \sum \sum_{t=1}^{N_{\rm sf}} R_{f,\ell}^t.  \label{eq:flowCon}
\eeqa
Observe that we have assumed that an arbitrary bandwidth fraction 
can be allocated.  Typical OFDMA systems such as LTE 
only permit scheduling of frequency-domain resources with the granularity of discrete subcarriers or groups of subcarriers (i.e. Resource Blocks).
 %
% After scheduling, these fractions can be rounded to the nearest integral number of subcarriers that is compatible with the numerology of the practical OFDMA system. However less optimal, this approach reduces the computational complexity of any optimizer since the continuous allocation can be solved for easily by convex optimization techniques, whereas discretized (i.e. subcarrier) BW allocation requires combinatorial integer assignment and is generally known to be a more difficult problem. Also it is fair to assume that little additional gain in utility and rate are sacrificed from rounding in this way when the number of subcarriers is large.

\section{Proposed Algorithm}
\label{sec:algorithm}

The problem as formulated is a binary mixed-integer optimization, which 
is non-convex.  A number of algorithms and heuristics have been proposed for OFDMA-TDD scheduling, many of which employ mixed-integer methods to optimize over the integral search space formed by the subcarrier and time slot indices \cite{AlRawi:11,Ahmed:12,ZhangX:13}. These techniques are designed to deal with the computationally hard problem posed by an inteference-limited network. As shall be demonstrated in the following section, two key aspects of our problem allow us to reduce the size of integral search space. Firstly, subframe allocation can be performed individually for each BS and RN without regard to interference at other nodes. Secondly, the hierarchical structure of the network is exploited to identify bottlenecks and strategically reassign resources to links to iteratively improve utility.

The suboptimal, greedy algorithm in \ref{alg:scheduler} operates by recursively updating slot configurations on each subtree. 
The procedure {\sc TestSubtree} takes as arguments, a root node $n$,
an initial TX/RX mode indicator matrix 
$\mathbf{X} = (x_i^t)$ for all  
nodes $i$ in the subtree rooted at $n$ and an optimal utility for that allocation. 
Note that $x_i^t$ represents whether a node can be transmitting ($x_i^t = +1$) or receiving ($x_i^t = -1$) in $t$, which differs from the notation introduced earlier for the link allocation $x_l$. In this scheme, the link allocation $x_l$ is determined by whether one adjacent node is transmitting while the other is receiving.
\beqa
    & x_\ell^t = 1, \forall \ell=(p,n) ~\leftrightarrow~ x_p^t = +1 \land x_n^t = -1, \label{eq:conHD3} \\
    & x_\ell^t = 1, \forall \ell=(n,p) ~\leftrightarrow~ x_p^t = -1 \land x_n^t = +1, \label{eq:conHD4}
\eeqa

The procedure returns a tuple ($\mathbf{X},U$) of the improved
duplexing schedule and utility for that subtree.
The procedure operates recursively, where at 
each level of the tree, it attempts to allocate additional subframes 
in either the DL and UL (by switching the TX/RX mode of nodes at that level) and calling the inner optimizer, here denoted as $\mathrm{Opt}()$, to test if utility has improved.
The inner optimizer utilizes the fact that, for any candidate duplexing
schedule, the optimization problem is the maximization of a concave function
with linear constraints.

We initially call $\mathrm{TestSubtree}()$ with the root node $n = BS$ to perform the following breadth-first procedure on the entire tree. In lines 3-10, we make a recursive call to reschedule each RN subtree with the initial $\mathbf{X}$. This tests the case where the current configuration is optimal for links adjacent to the BS but possibly not for subtrees rooted at a RN. We then reschedule a UL subframe in the DL (i.e. $x_{new} = +1$) at the BS by calling $\mathrm{Realloc}$ in line 13. We naturally set each UE in the set of children $C(n)$ of the root node (the BS, in this case) to RX mode in the same $t$; otherwise, these links would be unutilized in this subframe. When $\mathrm{Realloc}$ is called on a RN, it selects the first $t$ for rescheduling such that its transmission mode is the same as its parent. If no such $t$ exists, it simply selects the next subframe in succession.  The inner optimizer is then called, which returns the intermediate utility $U_{tst}$. While $U_{tst}$ may be less than the initial utility $U$ at this point, there is still the opportunity for it to improve by rescheduling subframes in each relay subtree (lines 18-29) (therefore, the procedure inside the \textit{while} loop continues as long as the intermediate utility $U_{tst}$ is increasing). In the base case, where the base of the tree has been reached or we have returned from testing all subtrees, we then attempt to improve utility by rescheduling slots in the UL using the same recursive procedure.

\begin{algorithm}
\caption{Dynamic TDD scheduler}
\label{alg:scheduler}
\begin{algorithmic}[1]
%\Procedure {Schedule} {$\mathbf{X}_{init}$}
%\State {$U \gets \Call{Opt}{\mathbf{X^\prime}}$}
%\State {$(U_{tst},\mathbf{X}_{tst}) \gets \Call{TestSubtree}{n_{BS}, \mathbf{X}, U_{tst}$}}
%\EndProcedure
%\newline
\Procedure {TestSubtree}{$n$, $\mathbf{X}$, $U$}
\State $U_{tst} \gets \infty$
%\State $\overline{R}_{f,opt} = \mathbf{0}$
\For {$r \in C(n) \vert r \in \mathcal{RN}$}
	\State {$(U_{tst}^{\prime},\mathbf{X}_{tst}) \gets \Call{TestSubtree}{r, \mathbf{X}, U$}}
	\If {$U_{tst}^{\prime} > U$}
		\State $U_{tst} \gets U_{tst}^{\prime}$
		\State $U \gets U_{tst}$
		\State $\mathbf{X} \gets \mathbf{X}_{tst}$
	\EndIf
\EndFor
\For {$x_{new} \in \{+1,-1\}$}
	\While {$U_{tst} > U$}
		\State $\mathbf{X}_{tst} \gets $\Call{Realloc}{$\mathbf{X}$,$n$,$x_{new}$}
		\For {\{$u \in C(n) \vert u \in \mathcal{UE}$\}}
			%\State $\mathcal{R}^\prime = \emptyset$ \Comment{Set children that are RNs}
			\State $\mathbf{X}_{tst} \gets $\Call{Realloc}{$\mathbf{X}_{tst}$,$u$,$-x_{new}$}
		\EndFor
		\State $U_{tst} \gets \Call{Opt}{\mathbf{X}_{tst}}$
		\For {\{$r \in C(n) \vert r \in \mathcal{RN}$\}}
			\State $\mathbf{X}_{tst} \gets $\Call{Realloc}{$\mathbf{X}$,$r$,$-x_{new}$}
			\State {$(U_{tst}^{\prime},\mathbf{X}_{tst}^{\prime}) \gets \Call{TestSubtree}{r, \mathbf{X}_{tst}, U$}}
			\If {$U_{tst}^{\prime} > U_{tst}$}
				\State $U_{tst} \gets U_{tst}^{\prime}$
				\State $\mathbf{X}_{tst} \gets \mathbf{X}_{tst}^{\prime}$
			\EndIf
		\EndFor
		\If {$U_{tst} > U$}
			\State $U \gets U_{tst}$
			\State $\mathbf{X} \gets \mathbf{X}_{tst}$
		\EndIf
	\EndWhile
\EndFor
\State \Return {($\mathbf{X}$, $U$)}
\EndProcedure
\end{algorithmic}
\end{algorithm}

%\State $(U_{opt}, \overline{R}_f) \gets \Call{Opt}{\mathbf{X}}$ \Comment{Init. call to "inner" optimizer}
%\While {$\lvert \mathcal{F} - \mathcal{F}_{done} \rvert > 0$} \Comment Some flows not done
%	\State $k_{worst} \gets \arg\min_k \overline{R}_{f_k}$ \Comment{Index of worst flow}
%	\If {$k(\bmod{2}) \equiv 0$}
%		$dir =$ 'DL' \Comment{DL flow}
%	\Else
%		~$dir =$ 'UL' \Comment{UL flow}
%	\EndIf
%	\State $p = parent(UE_i)$ \Comment{Parent node}
%	\If {$UE_i$ in outage $\mathbf{or} ~p \in \mathcal{T}^{dir}_{done}$}
%		\State {add $f_k$ to $\mathcal{F}_{done}$} \Comment{Exclude flow}
%		\State {$\mathbf{continue}$}
%	\EndIf
%	\State $p^\prime \gets p$
%	\While {$R_{p^{\prime,{dir}}} < R^{max}_{p^{\prime,{dir}}}$} \Comment{Bottleneck detection}
%		\State {$p^\prime \gets parent(p^\prime)$}
%	\EndWhile
%	\State {$\mathbf{X}^\prime = \Call{realloc}{\mathbf{X},p^\prime,dir}$} \Comment{Reassign SF in \textit{dir} at $p$}
%	\State {$(U^\prime, \overline{R}^\prime_f) \gets \Call{Opt}{\mathbf{X^\prime}}$}
%	\If {$U^\prime > U_{opt}$}
%		\State $U_{opt} = U^\prime$
%		\State $\overline{R}_f = \overline{R}^\prime_f$
%		\State $\mathbf{X} = \mathbf{X}^\prime$
%	\Else
%		\State {add $f_k$ to $\mathcal{F}_{done}$}
%		\State {add $p$ to $\mathcal{T}^{dir}_{done}$}
%	\EndIf
%\EndWhile

\section{Simulation Methodology \& Results}
\label{sec:sim_setup}
% need to justify omnidirectional received power interference model

% single-stream gain reduction (i.e.) bfGainRedStream

% using only 8 subframes instead of 10 since we neglect "special" subframes

% OFDM numerology from Samsung

Apart from the parameters taken from \cite{AkdenizCapacity:14}, we base our simulations on the 3GPP urban microcell (UMi) model and guidelines laid down by the 3GPP and ITU-R for modeling LTE/LTE-A networks \cite{3GPP36.814,ITU-M.2134}. The use of 1ms subframes is based on the Samsung mmW design, presented in \cite{PiSysDes:11}.

\begin{table}[h]\footnotesize
\centering
\caption{Network model parameters}
\label{tab:sim_params}
\begin{tabular}{ |p{0.01\columnwidth}|p{0.22\columnwidth}|p{0.55\columnwidth}| }
\hline
& \textbf{Parameter} & \textbf{Value} \\ \hline
\multirow{3}{*}{\rotatebox[origin=c]{90}{Antenna/RF~~~}}
& TX power & $P_{TX}=30 dBm$ (BS and RN)\newline$P_{TX}=20 dBm$ (UE) \\ \cline{2-3}
& Antenna & $n_{ant} = 64$, 8x8 $\lambda / 2$ planar array (BS and RN)\newline$n_{ant} = 16$, 4x4 $\lambda / 2$ planar array (UE) \\ \cline{2-3}
& Noise figure & $NF_{TX}=5$ (BS and RN)\newline$NF_{TX}=7$ (UE) \\ \cline{2-3}
\hline
\multirow{3}{*}{\rotatebox[origin=c]{90}{Channel~~~~~}}
& Bandwidth & $W=1000$ MHz \\ \cline{2-3}
& Carrier freq. & $f_c = 28$ GHz \\ \cline{2-3}
& Path loss &$\alpha = 72.0$, $\beta = 2.92$, $\sigma = 8.7dB$ (NLOS)\newline$\alpha=61.4$, $\beta=2$, $\sigma=5.8dB$ (LOS) \cite{AkdenizCapacity:14}\\ \cline{2-3}
& LOS-NLOS-outage prob.  & $a_{out}=0.0334m^{-1}$, $b_{out} = 5.2$, $a_{los} = 0.0149m^{-1}$ \cite{AkdenizCapacity:14} \\ \cline{2-3}
\hline
\multirow{3}{*}{\rotatebox[origin=c]{90}{TDD~~~}}
& Frame period & $T_f = 10ms$ \\ \cline{2-3}
& Subframe period & $T_s = 1ms$ \\ \cline{2-3}
& Subframe\newline allocation & Case 1: Static with reserved BS $\leftrightarrow$ RN SFs \newline Case 2: Dynamic \\ \cline{2-3}
\hline
\multirow{3}{*}{\rotatebox[origin=c]{90}{General~~~}}
& Area & $400m^2$ (wrap-around distance calculation)\\ \cline{2-3}
& Inter-site Distance (ISD) & RNs placed at $>50m$ from BS\\ \cline{2-3}
& Number of nodes & Case 1: $n_{BS}=1$, $n_{RN}=2$,$n_{UE}=10$\newline Case 2: $n_{BS}=1$, $n_{RN}=4$,$n_{UE}=10$ \\ \cline{2-3}
& Traffic & Full buffer \\ \cline{2-3}
& Number of drops & $N_{drop} = 25$ \\ \cline{2-3}
\hline

%FTP with file sizes: \newline $ R_f^{DL,max} = \{1, 2, 5, 10, 50, 100\}$ ~Mb \newline
%$ R_f^{UL,max} = \{0.5, 1, 2, 5, 25, 50\}$ ~Mb

\end{tabular}
\end{table}

We design our simulation scenarios to compare the performance of our dynamic TDD algorithm with the static TDD and relay configurations supported by LTE. For the static case, we test every valid combination of the TDD patterns in table 4.2-2 of \cite{3GPP36.211} and the possible reserved eNB-RN transmission subframes in table 5.2-2 of \cite{3GPP36216}. For each static configuration, we run the inner OFDMA optimization to provide a fair comparison with the dynamic scheduler.

We test a random network topology with 2 and 4 relay nodes. For each scenario and topology, a random drop is performed where nodes are placed uniformly randomly in the area, with the constraint that RNs must be over 50m from the wired BS. Note also that for RN-to-BS links, we assume that the deployment of RNs is planned by the operator to ensure a LOS backhaul connection.

%As explained in section \ref{sec:system_model}, the probability that a UE has a LOS or NLOS link or is in outage comes from the statistical model presented in \cite{AkdenizCapacity:14}. Certainly, we may see more unplanned deployments of future mmW cellular networks with self-organizing RNs that have LOS as well as NLOS backhaul links of varying quality. However the optimal cell selection and routing algorithm would need to be robust to such relaxed constraints on the topology, which is outside our problem scope.

%We apply a FTP traffic model to the network, which is similar to the 3GPP model in \cite{3GPP36.814}. Larger file sizes are, however, used to better demonstrate the capacity offered by mmW relaying. File sizes are randomly (uniformly) drawn from the sets in table~\ref{tab:sim_params} that limit the maximum total rate of a specific DL and UL flow.

\begin{table}[ht!] \footnotesize
	\caption{Mean fractions of LOS, NLOS, and outage UEs}
\label{table:outage_stats}
\centering
\begin{tabular}{ |p{0.15\columnwidth}|p{0.1\columnwidth}|p{0.1\columnwidth}|p{0.1\columnwidth}| }
\hline
 & \textbf{LOS} & \textbf{NLOS} & \textbf{Outage} \\ \hline
2RN 10UE & 0.388 & 0.600 & 0.012 \\ 
\hline
4RN 10UE & 0.536 & 0.460 & 0.004 \\
\hline
\end{tabular}
\end{table}

\begin{table}[ht!] \footnotesize
\caption{UE flow mean, median and cell edge (worst 5\%) rate for dynamic and static TDD (units in Mbps)} 
\label{table:perf}
\centering
\begin{tabular}{p{0.08\columnwidth}p{0.08\columnwidth}|p{0.08\columnwidth}|p{0.08\columnwidth}|p{0.07\columnwidth}|p{0.07\columnwidth}|p{0.07\columnwidth}|p{0.07\columnwidth}|p{0.07\columnwidth}}
\cline{3-8}
 & & \multicolumn{2}{c|}{\bfseries Mean} & \multicolumn{2}{c|}{\bfseries Median} & \multicolumn{2}{c|}{\bfseries  Cell edge} \\ \cline{3-8}
&  & DL & UL & DL & UL & DL & UL \\ \cline{1-8}
\multicolumn{1}{ |p{0.08\columnwidth}  }{\multirow{3}{*}{2RN} } &
\multicolumn{1}{ |c| }{DTDD}  & 31.73 & 26.65 & 20.86 & 10.76 & 0.35 & 0.19 &     \\ \cline{2-8}
\multicolumn{1}{ |c  }{}                        &
\multicolumn{1}{ |c| }{static}  & 21.20 & 8.80 & 7.70 & 0.63 & 0.23 & 0.15 &     \\ \cline{2-8}
\multicolumn{1}{ |c  }{}                        &
\multicolumn{1}{ |c| }{gain} & \textbf{1.50x} & \textbf{3.03x} & \textbf{2.71x} & \textbf{17.24x} & \textbf{1.57x} & \textbf{1.28x} &     \\ \cline{1-8}
\multicolumn{1}{ |p{0.08\columnwidth}  }{\multirow{3}{*}{4RN} } &
\multicolumn{1}{ |c| }{DTDD}  & 48.58 & 51.04 & 39.45 & 28.27 &  0.68 & 0.37 & \\ \cline{2-8}
\multicolumn{1}{ |c  }{}                        &
\multicolumn{1}{ |c| }{static}  & 29.61 & 11.32 & 8.34 & 1.11 & 0.32 & 0.16 \\ \cline{2-8}
\multicolumn{1}{ |c  }{}                        &
\multicolumn{1}{ |c| }{gain} & \textbf{1.64x} & \textbf{4.51x} & \textbf{4.73x} & \textbf{25.45x}  & \textbf{2.14x} & \textbf{2.36x} &     \\ \cline{1-8}
\end{tabular}
\end{table}

%\begin{figure}[ht!]
%\centering
%\includegraphics[width=0.8\columnwidth]{img/dtdd_vs_opt_rate.eps}
%\caption{CDF of DL and UL flow rates for DTDD and optimal allocations over 4 subframes}
%\label{fig:alg_perf}
%\end{figure}

\begin{figure}
\begin{subfigure}[ht!]{1.0\linewidth}
\centering
\includegraphics[width=0.8\columnwidth]{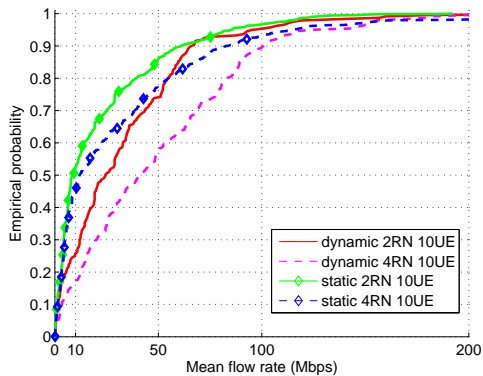}
\caption{Downlink}
\label{fig:dl_flow_rate_cdf}
\end{subfigure}

\begin{subfigure}[ht!]{1.0\linewidth}
\centering
\includegraphics[width=0.8\columnwidth]{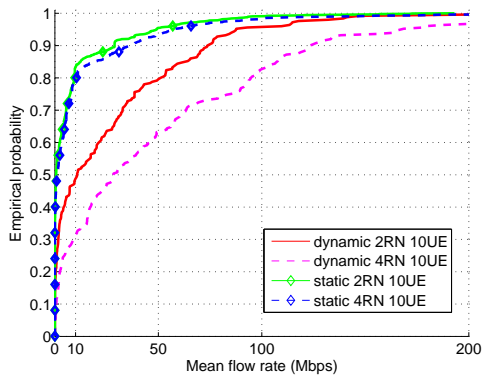}
\caption{Uplink}
\label{fig:ul_flow_rate_cdf}
\end{subfigure}
\caption{CDF of flow rates}
\end{figure}

The marked gains offered by optimized DTDD scheduling over static LTE TDD relaying are shown in Figures \ref{fig:dl_flow_rate_cdf} and \ref{fig:ul_flow_rate_cdf}. Also from table \ref{table:perf} we see how DTDD provides gains of over 1.5x in average downlink rate for the 2-relay and 4-relay cases, respectively. We find an even more distinct improvement in average uplink flow rate: 3x and 4.5x for the 2- and 4-RN scenarios. Improvement in median rates are dramatic, particularly for the uplink case, which further underscores the poor performance of the statically-configured LTE relays for uplink traffic. Edge rates, which we define as the lowest 5\% of rates, also improve notably in all cases. Under proportional fairness, the much lower rates of median and edge users in the static relay case result in large increases in system utility when DTDD allocation is used.
%(although utility is hard to quantify in any significant way to due being an essentially unitless metric). 

\begin{figure}[ht!]
\centering
\includegraphics[width=0.8\columnwidth]{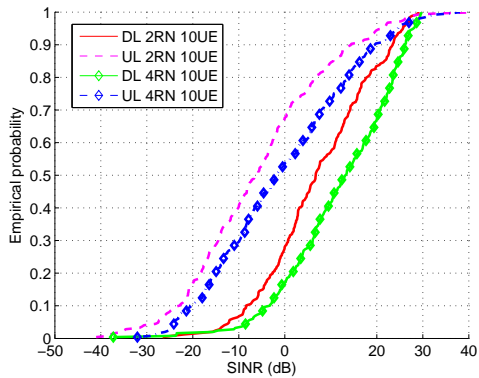}
\caption{CDF of DL and UL SINR for non-outage UEs after beamforming with omnidirectional interference}
\label{fig:sinr_cdf}
\end{figure}

We also find that, despite adding 4.76 dB of SINR on average, the gains of the 4-relay over the 2-relay network are only slight for the static TDD case, whereas they are much more pronounced (linear for the UL and nearly linear for the DL mean rates) under DTDD. This result highlights how adding relays provides little benefit when the eNB-RN links are bottlenecked, which is evidently the norm for LTE-style relays in dense mmW networks like we have simulated.

It is important to note that the results shown for the static LTE-relay case represent the average performance over all TDD schemes with 8 usable subframes. This does not illustrate the point that the network must initially select a single configuration, which may result in the ratio of uplink to downlink subframes and eNB-to-RN and RN-to-UE subframes being very poorly matched to the given the network loading. User data traffic in cellular networks is highly variable, so by this fact alone, the gains of DTDD should be even more significant than shown by these simulations. Also since different relays may have traffic loads that vary significantly, a single global frame configuration will likely not be appropriate for each relay and its users. In future work, we intend to demonstrate the full benefits of our scheme by developing a network model with time-varying channels and traffic.

\paragraph*{Performance and Optimality}
From comparing the performance of our suboptimal DTDD algorithm to the optimal allocation, which is computed via brute-force search, we find that for the 2-RN/10-UE case, DTDD achieves 13\% and 7\% off the optimal DL and UL mean flow rate and is within 5\% of the optimal utility. Our heuristic took 12 iterations (of calling the inner optimizer), on-average, and a maximum of 19 iterations to complete, as opposed to 329 for brute-force (found by numerically generating and counting unique permutations). We separately perform these simulations over only 4 subframes since the set of unique configurations we must search over increases exponentially with the number of subframes and would be intractable for 8 subframes. Clearly, the number of permutations of $N_{sf}$ subframes for a single BS and $N_{RN}$ relays is upper-bounded by $2^{N_{sf}(N_{RN}+1)}$, and while the number of unique permutations is much fewer, it still demonstrates that a brute-force approach is computationally unmanageable.

\section{Conclusions}

Multi-hop relaying provides an attractive technology for
extending coverage in mmW systems which are inherently
limited by range and blocking. In conventional cellular systems, the full benefits of relays are difficult to realize due
to the need to have global, common UL-DL time frame
assignments with reserved subframes for relays. However,
due to directional isolation with high-gain antennas, we have
argued that duplexing schedules in mmW systems can be
selected individually on each link. To exploit this flexibility,
we have proposed a dynamic duplexing TDD system, where
duplex schedules can adapt to local channel and traffic conditions. We
have formulated the selection of the duplex schedules as a
joint optimization with the bandwidth allocation to maximize
a global utility. The resulting optimization is non-convex,
but we have developed a computationally efficient suboptimal
algorithm based on recursive searching within subtrees. Simulations of realistic deployments demonstrate
considerable gains over static allocations, even under fairly
uniform traffic.

\bibliographystyle{IEEEtran}
\bibliography{bibl}

\end{document}